\documentclass[preprint2]{aastex}
\usepackage{natbib}

\newcommand{\Eq}   [1] {Eq.\ \ref{#1}}
\newcommand{\Eqn}  [1] {Equation \ref{#1}}
\newcommand{\Eqs}  [2] {Eqs.\ \ref{#1}--\ref{#2}}
\newcommand{\Fig}  [1] {Fig.~\ref{f:#1}}

\newcommand{\be}   [1] {\begin{equation}\label{#1}}
\newcommand{\ee}       {\end{equation}}
\newcommand{\bea}  [1] {\begin{eqnarray}\label{#1}}
\newcommand{\eea}      {\end{eqnarray}}

\newcommand{\av}[1]{\bra #1 \ket}

\newcommand{\half}{{\frac{1}{2}}}
\newcommand{\bra}{\langle}
\newcommand{\ket}{\rangle}

\newcommand{\avens}[1]{{\bra #1 \ket}_{\rm ens}}
\newcommand{\avxy}[1]{{\bra #1 \ket}_{xy}}
\newcommand{\avt}[1]{{\bra #1 \ket}_{t}}
\newcommand{\aver}[1]{\bar{#1}}

\newcommand{\fl}[1]{{\breve{#1}}}
\newcommand{\flux}[1]{{#1}}
\newcommand{\pp}[2]{{\partial#1\over\partial#2}}
\newcommand{\pdz}[1]{\pp{#1}{z}}

\newcommand{\Cw}{C(\omega,\Dt)}
\newcommand{\DDt}[1]{{D \over Dt} (#1)}
\newcommand{\DDtt}{{D \over {D t}}}
\newcommand{\Dnu}{\Delta\nu}
\newcommand{\Dt}{\Delta t}
\newcommand{\Em}{E_{\omega}}
\newcommand{\FFrad}{{\bf F}_{\rm rad}}
\newcommand{\FFv}{{\bf F}_{\rm visc}}
\newcommand{\Fcbar}{{\bar F_{\rm conv}}}
\newcommand{\Fkbar}{{\bar F_{\rm kin}}}
\newcommand{\Frad}{{\bar F}_{\rm rad}}
\newcommand{\Fvbar}{{\bar F}_{\rm visc}}
\newcommand{\Fvi}{F_{{\rm visc},i}}
\newcommand{\Gone}{\, \Gamma_1}
\renewcommand{\Im}[1]{\:{\rm Im}[#1]\:}
\newcommand{\Mm}{M_{\omega}}
\newcommand{\Pbar}{\bar{P}}
\newcommand{\Pgbar}{{\bar{P}_{\rm g}}}
\newcommand{\Pg}{P_{\rm g}}

\newcommand{\Ptbar}{{\bar P_{\rm t}}}

\newcommand{\Qdiss}{Q_{\rm diss}}
\renewcommand{\Re}[1]{\:{\rm Re}[#1]\:}
\newcommand{\Sij}{s_{ij}}
\newcommand{\Vm}{V_{\omega}}
\newcommand{\Wm}{W_{\omega}}
\newcommand{\dPg}{\delta \Pgbar}

\newcommand{\dPra}{{\dPr}^a}
\newcommand{\dPrn}{{\dPr}^n}
\newcommand{\dPr}{\dP_r}
\newcommand{\dPt}{\delta \Ptbar}
\newcommand{\dPwa}{{\dPw}^a}
\newcommand{\dPwn}{{\dPw}^n}
\newcommand{\dPws}{\dPw^{*}}
\newcommand{\dPw}{\dP_{\omega}}
\newcommand{\dP}{\delta \Pbar}
\newcommand{\ddt}[1]{\pp{}{t} (#1)}
\newcommand{\ddz}[1]{\pp{}{z} (#1)}
\newcommand{\ddzz}{\pp{}{z}}
\newcommand{\Div}{\nabla\!\cdot}
\newcommand{\dlnPra}{{\dlnP_r}^a}
\newcommand{\dlnPrn}{{\dlnP_r}^n}
\newcommand{\dlnPwa}{{\dlnP}_{\omega}^a}
\newcommand{\dlnPwn}{{\dlnP}_{\omega}^n}
\newcommand{\dlnP}{\delta \ln \Pbar}
\newcommand{\dlnr}{\, \delta \lnr}

\newcommand{\ek}{e_{\rm k}}

\newcommand{\eprm}{\fl{\et}}
\newcommand{\etbar}{{\bar\et}}
\newcommand{\ethbar}{\bar{e}_{\rm i}}
\newcommand{\etherm}{e_{\rm i}}
\newcommand{\et}{e}
\newcommand{\extra}[1]{\left[#1\right]}
\newcommand{\grad}{\nabla}
\newcommand{\lnr}{\ln\rb} 
\newcommand{\period}{.}
\newcommand{\pmode}{p-mode}
\newcommand{\rbar}{{\bar\rho}}

\newcommand{\rb}{\bar{\rho}}
\newcommand{\rprm}{\fl{\rho}}

\newcommand{\ubar}{\uzbar}
\newcommand{\uubar}{\bar{\uu}}
\newcommand{\uu}{{\mathbf{u}}}
\newcommand{\uzbar}{{\bar u_z}}

\newcommand{\uzp}{\fl{u}_z}

\newcommand{\uz}{\, \bar{u}_z} 
\newcommand{\vstress}{\sigma}
\newcommand{\vzzbar}{\bar{\vstress}_{zz}}
\newcommand{\xidz}{{\: {\partial \xid} \over {\partial z}}}
\newcommand{\xid}{\dot{\xi}}
\newcommand{\ximd}{\dot{\xi}_{\omega}}
\newcommand{\ximr}{{\partial\xim\over\partial r}}
\newcommand{\xim}{\xi_{\omega}}
\newcommand{\xirdz}{\: {{\partial \xird} \over {\partial z}}}
\newcommand{\xird}{\dot{\xir}}
\newcommand{\xirz}{\: {{\partial \xir} \over {\partial z}}}
\newcommand{\xir}{\xi_r}
\newcommand{\xiwdz}{\: {{\partial \xiwd} \over {\partial z}}}
\newcommand{\xiwd}{\dot{\xiw}}
\newcommand{\xiwz}{\: {{\partial \xiw} \over {\partial z}}}
\newcommand{\xiw}{\xi_{\omega}}
\newcommand{\xiz}{{\: {\partial \xi} \over {\partial z}}}

\begin{document}

\title{Solar Oscillations and Convection: I. Formalism for Radial
Oscillations}
\shorttitle{Solar Oscillations and Convection: I. Formalism}

\author{\AA. Nordlund}
\affil{Theoretical Astrophysics Center, and
    Astronomical Observatory, \\
    Juliane Maries Vej 30,
    2100 Copenhagen {\O}, Denmark}
\email{aake@astro.ku.dk}

\and

\author{R. F. Stein}
\affil{Dept. of Physics and Astronomy, Michigan State University,
East Lansing, MI 48823, U.S.A.}
\email{bob@steinr.pa.msu.edu}

\begin{abstract}

We present a formalism for investigating the interaction between
{\pmode} oscillations and convection by analyzing realistic,
three-dimensional simulations of the near-surface layers of the
solar convection zone.  By choosing suitable definitions for
fluctuations and averages, we obtain a separation that retains
exact equations.  The equations for
the horizontal averages contain one part that corresponds directly to the
wave equations for a 1-D medium, plus additional terms that arise
from the averaging and correspond to the turbulent pressure gradient
in the momentum equation and the divergence of the convective and
kinetic energy fluxes in the internal energy equation.  These terms
cannot be evaluated in closed form, but they may be measured in
numerical simulations.
The additional terms may cause the mode frequencies to shift,
relative to what would be obtained if only the terms corresponding
to a 1-D medium were retained---most straightforwardly by changing
the mean stratification, and more subtly by changing the effective
compressibility of the medium.  In the presence of time dependent
convection, the additional terms also have a stochastic time dependence,
that acts as a source of random excitation of the coherent modes.
In the present paper, we derive an expression for the excitation
power and test it by applying it to a numerical experiment of
sufficient duration for the excited modes to be spectrally resolved.
\end{abstract}


\section{Introduction}

The near-surface layers
of the Sun are of crucial importance for
the properties of the solar {\pmode} oscillations
(see the recent conference proceedings;
\citealt{GONG92,SOHO4,GONG94,BOMBAY96,SCORE96}).
The upper turning points of the {\pmode}s are located in these
layers and this is where the modes are excited and damped.
This is also where the solar convection zone gives
way to the visible solar photosphere.

The thin superadiabatic layer at the top of the solar convection zone is
characterized by large fluctuations in the thermodynamic variables.
As illustrated by detailed numerical simulations of the solar surface
layers
\citep{Nordlund82,%
Nordlund85f,%
Steffen+89,%
Nordlund+Dravins90,%
Nordlund+Stein91b,%
Steffen+Freytag91,%
Stein+Nordlund89b,%
Stein+Nordlund91b,%
Stein+Nordlund94a,%
Stein+Nordlund97a,%
Atroshchenko+Gadun94,%
Solanki+96},
the fluctuation amplitudes peak just below the visible surface,
where the temperature ranges from 5000 to 10,500 K, and the
logarithmic fluctuations of the density and pressure, $\Delta \ln \rho$
and $\Delta \ln P$,
are of the order of unity.
The velocity amplitudes are large throughout the photosphere, with rms Mach
numbers of the order of 0.3 and peak Mach numbers exceeding unity
in a small fraction of the volume
\citep{Nordlund+Stein91c,Stein+Nordlund97a}.

The layers with large amplitude fluctuations (at recent meetings referred
to as the `muck region') may be expected to influence the solar {\pmode}
oscillations in several ways.  First, these are the layers
where most of the random excitation of modes is expected to occur
\citep{Stein67,Stein68,Goldreich+Keeley77b,Goldreich+Kumar88,%
Goldreich+Kumar90,Stein+Nordlund91b,Bogdan+93,Goldreich+94b,Musielak+94}.
Second, the average vertical stratification of this region cannot
be assumed to be in hydrostatic equilibrium; the motions will
contribute an additional `turbulent pressure', that adds to the
normal gas pressure and hence tends to elevate the surface layers.
The large amplitude, non-linear fluctuations makes even the
definition of appropriate average values a non-trivial exercise.
Further, due to the extreme temperature sensitivity of the opacity
in the surface layers the emergent solar luminosity is produced by
an average state that differs noticeably from that of a
corresponding one dimensional model. Third, because of the presence
of the fluctuations,  the wave propagation properties of the medium
will in general be different than for a homogeneous medium.  We
follow \citet{Balmforth92b} and refer to the effects due to mean
structure changes as `extrinsic' (or `model') effects, and those
that are caused by changes in the wave propagation properties of
the medium as `intrinsic' (`modal' or `mode physics') effects.

A problem with reproducing the frequencies of the modes that have upper
turning points in these layers has indeed been known since the early
days of helioseismology
\citep{JCD+ea96science,
JCD88,
JCD+WD+YL88}
.
Improvements in the calculation of the equation of state
\citep{MHD90,OPAL96}
did not improve the situation,
but instead rather sharpened the significance of
the discrepancy between the observed and calculated oscillation
frequencies.

The discrepancy between the observed and theoretical mode frequencies
is primarily a function of frequency and is nearly
independent of degree, $\ell$, for small $\ell$.
It is small for the lowest frequencies, and grows to
significant values for frequencies approaching the cut-off frequency of
the solar photosphere (approximately 5 mHz).  This shows that the
cause of the discrepancy resides in layers to which the low frequency
modes hardly penetrate, but where the high frequency modes have a
significant amplitude.  Thus, the source of the discrepancy must lie near
the solar surface, in the outer layers of the solar cavity.

Stochastic excitation of {\pmode}s has been demonstrated in numerical
simulations of convection in the solar surface layers
\citep{%
Stein+88,
Steffen88b,%
Stein+Nordlund89a,
Stein+Nordlund91b,
Bogdan+93}.
In principle, the various contributions
to stochastic excitation, damping and frequency shift may be directly
measured in such numerical simulations. Alternatively, one may
instead extract information about the model structure and mode
propagation properties from the numerical simulations, and carry
that information over to standard envelope and mode calculation
procedures.  The advantage with the latter method is that one is
not limited to the studying the sparse spectrum of modes that are
excited in a small box.  The main purpose of the present paper is
to present a formalism for interpreting quantities used in such
1-D calculations as suitable averages of quantities that may be
measured in 3-D numerical simulations.

\citet{Stein+Nordlund91b} used a simplified version of the
formalism presented here in an initial study of mode excitation and
\citet{Rosenthal+97a,Rosenthal+98gong} also used a simplified
version to analyze the effect of 1-D/3-D model differences on the
frequencies of radial modes, showing that on the one hand the model
differences may account for the majority of the frequency
discrepancy, but that on the other hand the modal effects also are
significant.

In the present paper we develop a formalism for analyzing the
interaction of convection with purely radial oscillations by
choosing an exact decomposition into horizontal averages and
fluctuations.  Since most {\pmode}s are nearly radial near the solar
surface, the analysis covers the lower order behavior of non-radial
modes as well.

In Section \ref{sep.sec} we present the separation of variables
that we have chosen to work with.
In Section \ref{interaction.sec} we use this formalism to analyze
how the interaction of convection with the oscillations can cause mode
excitation, mode damping and frequency shifts, and in Section
\ref{example.sec} we show by explicit application to a numerical
experiment that the expression for the mode excitation produces
estimates of mode power that are consistent with what is actually
observed in the experiment.

In order to verify the formula for the mode excitation it is
necessary to use a numerical experiment of sufficiently long
duration for the excited modes to be spectrally resolved, because
this allows the mode excitation power to be estimated directly
from a measurement of the mode power and the mode lifetime
(obtainable from the full-width-at-half-maximum of the mode energy).
This necessitates a relatively low spatial resolution, which
precludes a direct comparison with the solar mode excitation power
(currently well established, see for example \citet{RocaCortes+99}).

In a subsequent paper (Paper II) we make use of simulations with
higher spatial resolution (and correspondingly shorter duration),
that allow us to make detailed comparisons with helioseismic data,
and to explore details of the processes that dominate the stochastic
mode excitation.

\section{Formalism}
\label{sep.sec}

Ideally, we would like to split the fluid equations into one set
describing the oscillations and one set describing convection.
By the very nature of the problem, such a separation cannot be
complete; if the convection is to excite the oscillations, it must
give rise to a source term in what would otherwise be equations
describing ideal, radial wave motions.  And if convection is to
have an effect on the (complex) frequencies of the modes, there
must be a possibility for convection to affect the compressibility
of the gas; in other words to have a coherent response with a
phase lag that in general may be expected to vary with height.
In addition, the convection is able to affect the frequencies
in a more trivial manner, namely by changing the overall
stratification relative to whatever reference 1-D model
one happens to compare with.

We are thus content with, and indeed looking for, a separation
consisting of wave-like equations with additional terms that are
related to the presence of convection.  Such a separation is
possibly not unique, but below we present one possibility.

One factor influencing our choice of separation was that
we are here concerned with a case that is, in a sense,
opposite to the more common case of small amplitude 3-D
fluctuations on top of large means.  We have large amplitude 3-D
fluctuations due to convection on top of a mean with small amplitude
coherent  {\pmode} fluctuations.
Thus we prefer to work with the actual equations,
rather than some truncated expansion.  We retain explicitly those
terms that cannot be worked out analytically, and subsequently measure
them in the numerical simulations.  In doing so, we wish to make
the split such that these terms have a well defined physical meaning,
and are numerically well conditioned.

\subsection{Notation and definitions}

To achieve the goals set out above, it is crucial to choose an
appropriate set of definitions for fluctuations and averages.
We use the convention that per-unit-volume quantities are written
in upper case and per-unit-mass quantities are written in lower case.

For the present case of purely radial motions,
we define straight horizontal averages
$\aver{F}$ and corresponding residuals $\fl{F}$
for per-unit-volume variables $F$,
\bea{1}
\aver{F}(z,t)       &=& \avxy{F} \\
\fl{F}(x,y,z,t)     &=& F(x,y,z,t) - \aver{F}(z,t) ,
\eea
and density weighted horizontal averages
$\aver{f}$ with residuals $\fl{f}$
for per-unit-mass variables $f$,
\bea{eq:maver}
\aver{f}(z,t)       &=& \avxy{\rho f}/\avxy{\rho} \\
\fl{f}(x,y,z,t)     &=& f(x,y,z,t) - \aver{f}(z,t) .
\eea
Thus, horizontal averages vanish for residuals
of per-unit-volume variables,
\be{3}
\avxy{\fl{F}} =  0 ,
\ee
while for the residuals
of per-unit-mass variables it is the
{\em mass weighted} horizontal averages
\be{5}
\avxy{\rho\fl{f}} = 0
\ee
that vanish.

Because of the large number of occurrences of horizontal averages,
we henceforth drop their explicit $xy$ subscript
($\avxy{} \rightarrow \av{}$) and retain subscripts only
on time averages ($\avt{}$) and ensemble averages ($\avens{}$).

As a consequence of \Eq{eq:maver},
products of two and three per-unit-mass variables obey
\bea{6}
\av{\rho f g} &=& \av{\rho \fl{f}\fl{g}}+ \rbar\bar{f}\bar{g} \\
\av{\rho f g h} &=&
\av{\rho \fl{f}\fl{g}\fl{h}}
+ \rbar\bar{f}\bar{g}\bar{h}
+ \nonumber \\ &&
  \av{\rho \fl{g}\fl{h}}\bar{f}
+ \av{\rho \fl{f}\fl{h}}\bar{g}
+ \av{\rho \fl{f}\fl{g}}\bar{h} \label{6a}
\eea

\subsection{1-D averaged equations}

To derive a set of hydrodynamic equations for the horizontal
averages, we apply the formalism of the
previous subsection to the three-dimensional equations of mass,
momentum and energy conservation,
\bea{8}
 \ddt{\rho} \!&=&\! - \nabla \cdot ( \rho \uu ) , \\
\label{8a}
 \ddt{\rho \uu} \!&=&\! - \nabla \cdot (\rho\uu\uu -{\bf \vstress})
 - \grad P - \rho {\bf g} ,\\
\label{8b}
\ddt{\rho\etherm} \!&=&\! - \nabla \cdot ( \rho \etherm \uu ) - P \, (\Div\uu)
\nonumber\\ &&
 - \Div \FFrad + \Qdiss
 \ ,
\eea
where $\etherm$ is the internal energy, $P=P(\rho,\etherm)$ is the gas
pressure, and $\FFrad$ is the radiative energy flux.
$\Qdiss$ is the viscous dissipation
\be{9b}
\Qdiss = \sum_{ij} \vstress_{ij} \Sij ,
\ee
where $\Sij$ is the symmetric part of the strain tensor
${\partial u_{i}/\partial x_{j}}$, and $\vstress_{ij}$ is the
viscous stress tensor, $\vstress_{ij} = \rho\nu\Sij$.

By \Eqs{6}{6a}, the total kinetic energy $\av{\half\rho u^2}$ splits
into the kinetic energy of the radial motion and of the horizontal
fluctuations,
\be{7a}
\av{\half\rho u^2} = \half\rbar\uzbar^2 + \av{\half\rho\fl{u}^2} ,
\ee
and the radial
kinetic energy flux $\av{\half\rho u^2 \uz}$ splits into
\bea{7b}
\av{\half\rho u^2\uz}
&=& \half\rbar\uzbar^2\uzbar + \av{\half\rho\fl{u}^2} \uzbar
\nonumber \\ {}
&+& \av{\rho\uzp^2} \uzbar + \av{\half\rho\fl{u}^2 \uzp} .
\eea
The terms on the RHS may be interpreted as the kinetic energy
flux of the radial motions, the advection of convective kinetic energy
density by the radial motions, the $P u$ flux associated with the turbulent
pressure (cf.\ below), and the radial kinetic energy flux associated
with the horizontal fluctuations.

An equation for conservation of total internal plus kinetic energy,
$\et = \etherm + \ek$, may be obtained
by adding the time derivative of the kinetic energy density,
$\ek = \half u^2$, to that for the internal energy $\etherm$,
\be{9}
 \ddt{ \rho \et} =
 - \nabla \cdot ( \rho\et \uu + P\uu  + \FFrad + \FFv)
 + \rho \uu \cdot {\bf g} ,
\ee
where $\FFv$ is the viscous flux
\be{9a}
\Fvi = - \sum_j u_j \vstress_{ij} .
\ee

Horizontal averaging of
the above equations yields equations that depend on
height  and time, but not on the horizontal coordinates:
\be{massav.eq}
 \ddt{\rbar} = - \ddz{\rbar\uzbar} ,
\ee
\bea{momav.eq}
 \ddt{\rbar \uzbar} &=& - \ddz{\rbar\uzbar^2 +\Pgbar -\vzzbar}
 + g \rbar \nonumber  \nonumber \\
 && \extra{- \ddz{\Ptbar}} ,
\eea
\bea{eintav.eq}
 \ddt{\rbar \ethbar}  \!&=&\!
 - \ddz{\rbar\ethbar \uzbar} - \Pgbar\pdz{\uzbar} \nonumber \\
 && [-  \pp{}{z}(\Fcbar + \Frad) \nonumber \\
 && {}+ \Qdiss + \av{\fl{\uu}\cdot\grad{P}}] ,
\eea
\bea{energyav.eq}
 \ddt{\rbar \etbar}  \!&=&\!
 - \ddz{\rbar\etbar \uzbar + \Pbar \uzbar} + g\rbar\uzbar \nonumber \\
 && [-  \pp{}{z}(\Fcbar + \Frad \nonumber \\
 && {}+ \Fkbar + \Fvbar ) ],
\eea
where $\rbar = \av{\rho} $, $\rbar\uzbar = \av{\rho
u_z} $, $\Pgbar = \av{\Pg}$, $\Pbar=\Pgbar+\Ptbar$,
and $\rbar\etbar = \av{\rho e}$.  $\Ptbar$ and $\Fcbar$
are defined below (\Eqs{14}{15}).

Equations \ref{massav.eq}--\ref{energyav.eq}
correspond closely to the hydrodynamic equations for a stratified,
homogeneous 1-D medium.
The terms that are not bracketed are exactly what appear in the 1-D
equations, whereas the extra, bracketed terms are due to
the convective motions.  These are the gradient of the turbulent
pressure
\be{14}
\Ptbar = \av{\rho\uzp^2} ,
\ee
the divergence of the convective flux
\be{15}
\Fcbar = \av{ (\rho\etherm + \Pg)\uzp }  ,
\ee
and the divergence of the
kinetic energy flux associated with convection,
\be{16}
\Fkbar  = \av{ (\half\rho u^2)\uzp } .
\ee
Here $\uzp$ is the vertical velocity relative to a frame of
reference moving with velocity $\uzbar = \av{ u_z\rho
}/\av{\rho}$; we  refer to this frame of reference
as `pseudo-lagrangian'.
Note that, in general, $\av{ \uzp } \neq 0$.

The $\Frad$ term should, in principle, not be bracketed, since
it also may appear in a 1-D medium.  It is, however, so intimately
related to the other flux terms that we prefer to place it inside
the brackets.  Furthermore, because of the strong non-linearities
involved, the value of $\Frad$ may differ substantially between 1-D
and 3-D models with the same average structure.

A stationary state that has $\uzbar=0$ obeys
\be{mommean.eq}
\ddz{\Pgbar + \Ptbar - \bar{\vstress}_{zz}} = g \rbar
\ee
\be{energymean.eq}
\ddz{ \Fcbar + \Fkbar + \Fvbar + \Frad} =  0
\ee

Note that the horizontally averaged pressure is in general not the same
as the pressure for the average density and energy of which it is a
function.  Because of the correlation of the fluctuations and the
contribution from higher order terms,
\be{17}
  \Pgbar = \Pg ( \rbar , \etbar )
  + {\partial^2 \Pg \over \partial \rho \partial e} \av{\rprm\eprm}
  + \mbox{higher order terms}
\ee
is in general not equal to $ \Pg ( \rbar , \etbar )$.

\subsection{Pseudo-Lagrangian 1-D Equations}

For the purpose of studying adiabatic (or near adiabatic) wave mode
fluctuations, it is more relevant to write the horizontally averaged
equations
in the pseudo-lagrangian reference frame (moving with velocity
$\uzbar$).  The operator
\be{18}
\DDt{} = \ddt{} + \uzbar \ddz{} ,
\ee
picks up the time variation in that frame of reference.  For a
per-unit-mass variable $f$ that satisfies
\be{19}
\ddt{\rho f} = -\Div{\flux{F}} ,
\ee
the change to a pseudo-lagrangian frame results in
\bea{20}
\DDt{\rho f} &=& -\Div{\flux{F}}+ \uzbar \ddz{\rho f} \nonumber\\
&=& -\Div{\left(\flux{F} - \rho f \uubar\right)} - \rho f \ddz{\uzbar} ,
\eea
i.e., the subtraction of the average flux of $\rho f$ from
within the divergence operator, and the addition of a
term of the form $-\rho f \ddz{\uzbar}$, that
accounts for the dilation or concentration of the (per unit
volume) quantity $\rho f$, corresponding to stretching or compression of the
coordinate system.

In terms of the $\DDt{}$ operator the averaged 1-D equations become
\bea{massavL.eq}
 \DDt{\rbar} &=& -\rbar\ddz{\uzbar} \\
\label{momavL.eq}
 \DDt{\rbar \uzbar} &=& - \ddz{\Pgbar-\vzzbar}
 + g \rbar
 - \rbar\uzbar\ddz{\uzbar}
 \nonumber \\ &&
 [- \ddz{\Ptbar}]
  ,\\
\label{energyavL.eq}
 \DDt{\rbar \etbar}  &=&
 - \ddz{\Pbar \uzbar} + g\rbar\uzbar
 - \rbar\etbar\ddz{\uzbar}
 \nonumber \\ &&
 [-  \ddz{ \Fcbar + \Fkbar + \Fvbar + \Frad }] .
 \nonumber \\ &&
\eea
The transformation between per-unit-volume and per-unit-mass
pseudo-lagrangian time derivatives is
\bea{22}
\rbar \DDt{\bar{f}} &=& \DDt{\rbar\bar{f}} - \bar{f}\DDt{\rbar} \nonumber\\
&=& \DDt{\rbar\bar{f}} + \rbar\bar{f} \ddz{\uzbar} .
\eea
The dilation term that is present in the per-unit-volume
formulation thus drops out in the per-unit-mass formulation,
because a fixed amount of mass is under consideration---the mass element
$\rbar dz$ is indeed suitable for depth integration of pseudo-lagrangian
per-unit-mass quantities.

In per-unit-mass variables, the equation of motion and the
energy equations for horizontal averages thus become
\bea{momavM.eq}
 \rbar \DDt{\uzbar} &=& - \ddz{\Pgbar+\Ptbar - \vzzbar}
 + g \rbar ,\\
\label{eintavM.eq}
 \rbar \DDt{\ethbar}  &=& - \Pgbar\pdz{\uzbar} \nonumber \\
 && [{}- \pp{}{z}(\Fcbar + \Frad) \nonumber \\
 && {}+ \Qdiss + \av{\fl{\uu}\cdot\grad{P}}] , \\
\label{energyavM.eq}
 \rbar \DDt{\etbar}  &=&
 - \ddz{\Pbar \uzbar}  + g\rbar\uzbar \nonumber \\
 && [{}- \ddz{\Fcbar + \Frad \nonumber \\
 && {} + \Fkbar + \Fvbar}] .
\eea
In what follows, we
use the traditional notation $\delta \Pgbar$, $\delta \rbar$, etc., to
distinguish pseudo-lagrangian perturbations from Eulerian ones.

\section{Interactions between oscillations and convection}
\label{interaction.sec}

We now use this formalism to calculate the work done on radial
oscillation modes by convection.  We then test the resulting expression
on the vertical resonant modes excited in a numerical simulation of
solar convection.

In the presence of a coherent mode, the time variation of the
additional, convective terms in \Eqs{massav.eq}{energyav.eq}
is partly in unison with the mode, reflecting the coherent response of
convection to the presence of the mode.  The coherent response may
again be divided into one part that is in phase with the mode
(appearing as the real part of a Fourier transform), and one part that
is in quadrature with the mode (the imaginary part of a Fourier
transform).  The imaginary part of the coherent component causes an
exponential damping (or growth) of a trapped mode, and the real part
causes a frequency shift.  In analogy with simpler situations we
also refer to these parts as ``adiabatic'' and ``non-adiabatic''.
There is also an incoherent contribution, corresponding to the random
variation of the convection, that contributes to these averages, and
produces stochastic mode excitation and damping.

The equation describing the time evolution of the mode
kinetic energy may be obtained from \Eq{momavM.eq}:
\bea{24}
\rbar \DDt{(\half\ubar^2)} &=&
- \ddzz [ \ubar ( \Pgbar + \Ptbar - \vzzbar)]
\nonumber \\ && {}
+ (\Pgbar + \Ptbar - \vzzbar) \pdz{\ubar}
\nonumber \\ && {}
+ \rbar \ubar g
\period
\eea
Integrating over time and depth, we obtain
\bea{25}
{\left[ \int dM \half \ubar^2 \right]}_0^t
&=& \int\! dt\!\!\int \! dz ~( \dPg + \dPt
\nonumber \\ &&
-\delta\vzzbar)\pdz{\ubar}
\nonumber \\ &&
+ \int\!dt\!\int \! dz ~ \rbar\ubar g
\nonumber \\ &&
+~[ \ ...\  ]_{\rm{boundaries}}
\period
\eea
where, $[ \ ...\  ]_{\rm{boundaries}}$ denotes the boundary contributions
that result from integrating the divergence form of the
equation.  For suitably chosen boundary conditions, these contributions
vanish.  Specifically, the displacement may be chosen to vanish at the
lower boundary, and the pressure may be chosen sufficiently small at the
upper boundary.  The work done by gravity vanishes if there are no net
mass displacements in the model.

The $\int \!dt\! \int \!dz~(\dPg+\dPt-\delta\vzzbar)\partial \ubar/\partial z$ part of the
work integral represents the PdV work done by the gas pressure,
the turbulent pressure, and the mean viscous stress.
(The mean viscous stress may be expected to be negligibly small
in stellar atmospheres and envelopes, and should be small also
in numerical simulations.)
The $\partial\ubar/\partial z$ factor is equal to
$-D\lnr /Dt = D \ln V / Dt$, where $V$ is the specific volume.
Pressure perturbations must
be out of phase with density perturbations in order to contribute to
the work integral.  In a diagram showing $\dP$ against
$\delta \rbar$ this corresponds to open curves, with a
counter clockwise sense of orientation.

The signs are as to be expected from basic physical principles;
the mode kinetic energy increases if the pressure is larger
during expansion than during compression.

\subsection{Coherent and incoherent fluctuations}
\label{coh-incoh.sec}

The PdV work integral, \Eq{25}, is
\be{w1}
  W = \int \!dt\! \int \!dz~ \dP \,\xidz .
\ee
Here $\dP$ is the pseudo-lagrangian total pressure fluctuation
(neglecting the small viscous stress contribution)
\be{w2}
\dP = \dPg + \dPt ,
\ee
and $\xi$ is the displacement, which is related to the velocity by
\be{w3}
\xi = \int^t dt' \uz ,
\ee
and to the density variations by
\be{w4}
{{D \lnr} \over {D t}} = - \xidz .
\ee
The displacement and the pressure fluctuations can be split into two
parts: one the coherent, sinusoidal, modal part and the other the
incoherent, random part:
\bea{w5}
\xi & = & \xiw + \xir\\
\dP & = & \dPw + \dPr .
\eea
The pressure fluctuations can be further split into an adiabatic part,
proportional to the density fluctuations,
\be{w6}
\delta \ln \Pbar^a = \Gone(z) \dlnr = - \Gone(z) \xiz,
\ee
and a non-adiabatic part, the remainder:
\bea{w7}
\dPw & = & - \Pbar \Gone(z) \xiwz + \dPwn , \\
\label{w7a}
\dPr & = & - \Pbar \Gone(z) \xirz + \dPrn ,
\eea
so that,
\be{w8}
\dP = \dPwa + \dPwn + \dPra + \dPrn .
\ee
Thus, the work integral can be expanded into the product of 4 pressure
fluctuation terms multiplied by 2 displacement terms,
\bea{w9}
\lefteqn{W = \int\int  dt dz}
\nonumber \\
& & \left[ \dPwa + \dPwn + \dPra + \dPrn \right]
    \left[ \xiwdz + \xirdz \right]
\nonumber \\
& & = \int\int  dt dz
\nonumber \\
& & \left[ \; (a) \; + \;  (b) \; + \; (c) \; + \; (d) \; \right]
    \left[ \; (1) \; + \; (2) \; \right] .
\eea
Consider the contribution of each possible pair to the work.
The integral as a whole may be expected to display a  ``random walk''
behavior; i.e., its expectation value grows with total time
of integration.  Contributions that are bounded are therefore
negligible.  The $(a) \times (1)$ contribution, for example,
\bea{w10}
(a) \times (1) \; &\propto& \; \xiwz \xiwdz
\\   &=& \DDtt \left[ \half \left(\xiwz\right)^2 \right]
\\  &\rightarrow& \half \left[ \left(\xiwz\right)^2 \right]_{t1}^{t2} ,
\eea
is a total integral, so it is negligible.
\bea{w11}
(c) \times (2) \; &\propto& \; \xirz \xirdz
\\  &=& \DDtt \left[ \half \left(\xirz\right)^2 \right]
\\  &\rightarrow& \half \left[ \left(\xirz\right)^2 \right]_{t1}^{t2} ,
\eea
so its contribution is also negligible.
\bea{w12}
\lefteqn{(a) \times (2) +(c) \times (1) \; \propto \;}
\nonumber \\ & &
  \Pbar \Gone \xiwz \xirdz + \Pbar \Gone \xirz \xiwdz =
\nonumber \\ & &
\Pbar \Gone \DDtt \left(\xiwz \xirz \right)
\nonumber \\ & &
  \rightarrow \Pbar \Gone \left[\xiwz \xirz \right]_{t1}^{t2} ,
\eea
and also gives a negligible contribution.  Note that this step
depends on using the
same $\Gone$ in the definitions of the adiabatic modal and random parts of
the pressure fluctuation (Eqs.\ \ref{w7} and \ref{w7a}).
\be{w13}
(b) \times (1) \; \propto \; \dPwn \xiwdz ,
\ee
represents the linear driving/damping of the mode.  It is the
balance of this term with the stochastic driving that determines
the amplitude of the mode.
\bea{w14}
(b) \times (2) \; &\propto& \; \dPwn \xirdz
\\  &\propto& \Pbar \dlnPwn \omega \Gone^{-1} \dlnPra ,
\eea
is a stochastic driving term.  However, it is small in comparison
to the next term, because it is
proportional to the non-adiabatic coherent pressure fluctuation which
is small since the mode amplitude growth time is many periods.
\bea{w15}
(c) \times (1) \; &\propto& \; \dPrn \xiwdz \\
    &\propto& \Pbar \dlnPrn \omega \Gone^{-1} \dlnPwa ,
\eea
is the dominant stochastic driving term.  It is proportional to the
non-adiabatic random pressure fluctuations which are large and the
adiabatic coherent pressure fluctuations which are also large, since
they provide the mode restoring force.
\be{w16}
(d) \times (2) \; \propto \; \dPrn \xirdz ,
\ee
does not represent any work on the mode (it does not contain any factor
representing the mode).

\subsection{Stochastic excitation}
\label{exc.sec}

Using the results from the previous subsection, we may derive
expressions that measure the stochastic excitation and the linear
damping in numerical simulations.  After proper scaling, such
measurements yield estimates of global excitation power and damping
that may be compared directly with estimates from observations
\citet{RocaCortes+99},
and with estimates from analytical theories (e.g.,
\citealt{Balmforth92c,Goldreich+94b}).

The mode energy per unit surface area (at $r = R$) is
\be{e1}
\Em
= \half \omega^2 \int_{r} dr ~ \xim^2 ~ \rho ~
\left({r \over R}\right)^2
\equiv \Mm \Vm^2
\ee
where $\Mm$ is (by definition) the mode mass, and
$\Vm = \ximd (R)$ is the mode velocity amplitude at the
reference radius $R$.
The change in a mode's kinetic energy over a time interval $\Dt$ is
\bea{e2}
\Delta \Em
&=& \int_{\Delta t} dt \int_{r} dr ~ \dP ~
  {\partial \ximd \over \partial r}
\nonumber\\
&\equiv& \int_{\Delta t} dt ~ \Em^{\half} ~ \Wm(t)
\nonumber\\
&\equiv& \Em^{\half} \: \Cw
,
\eea
where
\be{e3}
\Wm(t) \equiv \Em^{-\half} \int_{r} dr ~ \dP ~
{\partial \ximd \over \partial r}
,
\ee
is the instantaneous work integral for the given mode's displacement,
normalized with the square root of the mode's energy, and
\be{e4}
\Cw = \int_{\Dt} dt ~ \Wm(t)
\ee
is the work integrated over the time interval $\Dt$.

For small changes of amplitude $\Vm \rightarrow \Vm+\Delta\Vm$.
\Eqn{e1} with the definitions in \Eq{e2} give,
\be{e4a}
\Delta \Vm
= { \half \Mm^{-\half} \Cw }
.
\ee
For a particular realization of the stochastic driving, there is a
complex phase factor $e^{i\phi}$ between $\Delta\Vm$
and $\Vm$.  The ensemble average of the mode energy $\Em + \Delta \Em$
at $t+\Dt$ is
proportional to the ensemble average of $|\Vm + \Delta\Vm|^2$ over
all phases. In this averaging the linear term $\avens{\Vm\Delta\Vm}$
vanishes, and one obtains from the quadratic term
\bea{e6}
{\Delta \avens{|\Vm^2|} }
&=& \avens{ |\Delta  \Vm  |^2 } \\
&=& \avens{ |\half \Mm^{-\half} \Cw|^2 } ,
\eea
so that
\bea{e7}
{\Delta \avens{\Em} \over \Dt  }
&=& {\Mm \Delta \avens{|\Vm^2|} \over \Dt} \\
&=& {1 \over {4 \Dt} } \avens{|\, \Cw \,|^2}
.
\eea
In terms of Fourier transforms over the time interval $\Dt$, the
integrated work
\bea{e8}
 \Em^{\half} \Cw
&=& \Dt \Re{ \int_{r} dr ~ \dPws i\omega \ximr }
\nonumber\\
&=& \omega \Dt \Im{ \int_{r} dr ~\dPws \ximr }
\nonumber\\
&\equiv& \omega \Dt \Im{ \widehat{\Wm} }
\eea
is proportional to the
Fourier amplitude of the projection of the stochastic pressure
fluctuations onto $\xim$.

The expectation value of the square of the imaginary part
is half of the power of this random function in the frequency
interval $\Dnu = 1/\Dt$, since $\avens{| \widehat{\Wm} |^2} =
\avens{|\Re{\widehat{W}} |^2} + \avens{| \Im{\widehat{W}} |^2}$,
and the ensemble averages for the real and imaginary parts are
equal for a function with randomly distributed phases, so
\be{e9}
\avens{| \Im{\widehat{W}} |^2}
= \half \avens{| \widehat{\Wm} |^2}
.
\ee
From Eqs.\ \ref{e7}, \ref{e8} and \ref{e9}, we obtain
\bea{e10}
{\Delta \avens{\Em} \over \Dt  } & = &
  {1 \over {4 \Dt}} \avens{|\, {{\omega \Dt} \over \Em^{\half}}
    \Im{\widehat{\Wm} }\,|^2} \nonumber\\
  & = & {{\omega^2 \Dt} \over { 8 \Em }} \avens{|\, \widehat{\Wm}\,|^2}\nonumber\\
  & = & {{\omega^2 \Dt} \over { 8 \Em }} \avens{|\int_{r} dr ~ \dPws ~ \ximr |^2 }
,
\eea
and with the definition \Eq{e1} of the mode energy, the
expectation value of the stochastic excitation power is
\be{e11}
{\Delta \avens{\Em} \over \Dt  }
= { |\int_{r} dr ~ \dPws ~ \ximr |^2 \over 4 ~ \Dnu ~
\int_{r} dr ~ |\xim|^2 ~ \rho ~ ({r \over R})^2  } .
\ee
%
From \Eq{w13}, the damping rate is
\be{d1}
{1\over \Em} {d \Em \over dt}
= {\int_{r}dr ~ \dP ~ {\partial \ximd \over \partial r} \over
   \half \omega^2 \int_{r} dr ~ |\xim|^2 ~ \rho ~
   \left({r \over R}\right)^2 }
,
\ee
or, in terms of Fourier transforms,
\be{d2}
{1\over \Em} {d \Em \over dt}
= {2   \int_{r} dr ~ \Im{\dPws ~ \ximr}  \over
   \omega \int_{r} dr ~ |\xim|^2 ~ \rho ~
   \left({r \over R}\right)^2 }
.
\ee

\Eqn{e11} expresses the excitation power per unit surface
area in terms of the power density of the stochastic pressure
fluctuations, $\dPw$.  If these are measured in a numerical
simulation covering a small ($L \times L$) patch of the solar
surface, the result must be properly scaled in order to estimate
global excitation power.  Taking the results at face value would
correspond to assuming a periodic, coherent repetition of the
pressure fluctuations over the entire solar surface area $4\pi R^2$.
If the actual pressure fluctuations are instead uncorrelated over
scales larger than the scale of the numerical model, the global
spectral power density is smaller by a factor of $L^2/(4\pi
R^2)$.  The integrated excitation power is then obtained
by multiplying \Eq{e11} with the horizontal surface area of the
numerical model, $L^2$, rather than by the total surface area,
$4\pi R^2$.  (Note that this does not imply that the scale $L$
enters into the final result, since the expectation value of the
fluctuations of the mean pressure $\dPw$ drops correspondingly,
if the size of the numerical model is increased beyond the size
over which pressure fluctuations are correlated.)

\section{Examples}
\label{example.sec}
To demonstrate that the proposed formalism is useful, both from
a theoretical and a practical point of view, we provide two example
applications; a theoretical discussion of deviations from hydrostatic
equilibrium, and a practical application to a numerical data set.

\subsection{Deviations from hydrostatic equilibrium}

The equations for the horizontal averages (either \Eqs{massav.eq}
{energyav.eq} or the equivalent Lagrangian equations
\ref{massavL.eq}--\ref{energyavL.eq}) do not assume anything about
the horizontally averaged (barred) quantities, other than that
they are instantaneous, horizontal averages.

Assuming now that a stationary reference state exists, with
$\avt{\uz}=0$, one may ask what happens if initially the
barred variables deviate from the equilibrium state.
The initial state may then be taken as an initial condition
for a time integration of \Eqs{massavL.eq}{energyavL.eq}.

Ignoring in a first analysis the lagrangian time variation
of the terms that represent the convective flux, the radiative
flux, and the turbulent pressure, one has a set of equations
that are directly analogous to the ordinary, one dimensional
hydrodynamic equations, and admits exactly the same type of
solution; the initial non-equilibrium state slumps towards
the equilibrium state, overshoots, and a series of oscillations
ensues.  If the boundaries are closed, and there are no dissipative
terms, the oscillations will continue undamped, and consist
of a superposition of the eigenmodes of the system.

If the boundaries allow wave energy to escape, and / or there
are other dissipative effects, the eigenmodes will be damped, and
the solution will eventually settle to the equilibrium solution.

Likewise, if one considers the actual, time-dependent lagrangian
response of the convective and radiative fluxes, and of the turbulent
pressure, these will in general also contribute to the damping (or
possibly self-excitation) of the eigenmodes of the system.

In conclusion, an initial state that is not in hydrostatic
equilibrium may be thought of as a superposition of (in general
damped) eigenmodes, and such a system will typically perform
damped oscillations around a stationary reference state.

All of this might be analyzed by making use of the formalism put forward
here, at any chosen level of realism; one may replace the bracketed
terms that are not available in closed form with theoretical estimates,
or one may retain them as they are, and evaluate them from numerical
simulations.  In fact, we have performed numerical experiments of just
this type by subjecting snap shots from 3-D simulations to large
amplitude perturbations of the vertical equilibrium, in order to study
in particular the damping properties of the resulting set of
oscillations.  Results of the analysis will be published in a
forthcoming paper.

\subsection{Excitation rates and energy losses in a numerical experiment}

As a direct test of the expression for the excitation rate, \Eq{e10},
we have evaluated the expression numerically, using data from a numerical
experiment covering 43 hours of solar time at a resolution of
$63\times63\times63$---additional details about this experiment are
given elsewhere.  The length of this run is sufficient for
a direct measurement of mode line widths, and hence for a direct
determination of the right hand side of the expression
\be{excbalance}
{\Delta \avens{\Em} \over \Dt} =
{E_{\rm mode} \over \tau_{\rm mode}} ,
\ee
where the mode energy life time $\tau_{\rm mode}$ is related to the
full width at half maximum of the mode energy spectrum
$\Delta\nu_{\rm FWHM}$ by
\be{FWHM}
2 \pi \tau_{\rm mode}\Delta\nu_{\rm FWHM} = 1 .
\ee
\begin{figure}
\plotone{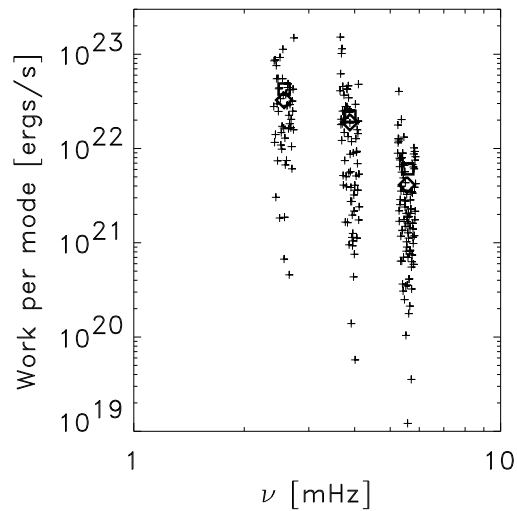}
\caption[]{The excitation power (small pluses) in the
neighborhood of the frequencies of the three radial modes that
are excited to measurable amplitudes in the box.  The average
excitation is shown as diamonds, and the average energy loss per
unit time is shown as squares.  The results are from a numerical
experiment with realistic opacities and equation of state, covering
43 hours of solar time at a resolution of $63\times63\times63$
\citep{Georgobiani+99b}.
Note that, because of the relatively low resolution required
for such a long run to be affordable, the results are not directly
comparable to solar values.  The mode mass of the box also differs
significantly from the solar mode mass at the same frequency.}
\label{f:work.ps}
\end{figure}
\begin{figure}
\plotone{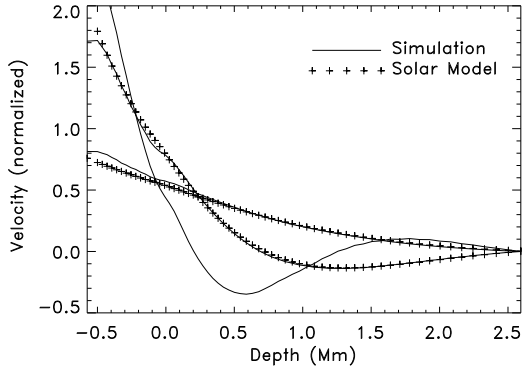}
\caption[]{The velocity mode shapes for the three
radial oscillations, at 2.5, 3.9, and 5.6 mHz, that are observed in
the 43 hour simulation.  For comparison, the corresponding modes
from Model S of \citet{JCD+96science}, normalized to the same mode
mass within the box, are also shown (plus symbols) for the two modes
whose frequencies are below the acoustic cut-off frequency.}
\label{f:modes.ps}
\end{figure}
\begin{figure}
\plotone{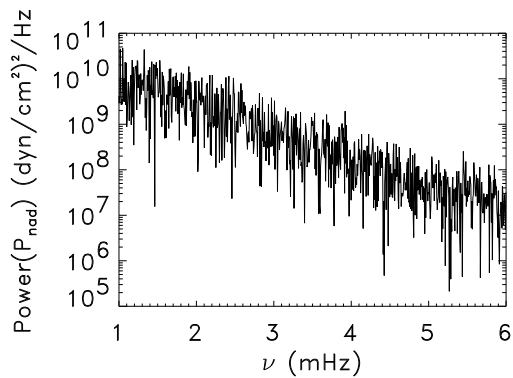}
\caption[]{The power-spectrum of the non-adiabatic pressure
fluctuations, \Eq{w7a}, in a layer close to the surface of the 43 hour
simulation.}
\label{f:Pna.ps}
\end{figure}
In \Fig{work.ps} we compare estimates of the excitation power
based on \Eq{e11} with the actual energy loss
for the three radial modes that are observed in the experiment.
The energy loss of each mode was computed from \Eq{excbalance}
by fitting Lorentzians
to the excess mode power above the background noise (cf.\ Paper II,
Fig.\ 1), using $E_{\rm mode}$ and $\tau_{\rm mode}$ as parameters
in the fit.  The excitation power was evaluated from \Eq{e11},
with mode displacement factors $\ximr$, and $\xim$, obtained by
using Fourier transforms to project out the three modes 
(cf.\ \Fig{modes.ps}).  Note that the actual amplitudes of
these modes do not enter into the estimate of the excitation power,
because of the normalization with the mode kinetic energy in \Eq{e11}.
The non-adiabatic pressure fluctuations at the mode frequencies,
$\dPws$, were obtained from the Fourier transform of the non-adiabatic
pressure fluctuations (\Fig{Pna.ps}), averaging over a neighborhood
of the mode frequencies to obtain the values indicated with diamonds 
in \Fig{work.ps}.
For further details on how the formulae are evaluated see the appendix
of Paper II.

Note that the spectrum
of non-adiabatic pressure follows a power law and shows no
particular features at the resonant mode frequencies.  This illustrates
that the non-adiabatic fluctuations are mostly incoherent.
Conversely, since the modes are only weakly damped, the coherent
fluctuations at the mode frequencies are mostly adiabatic. At
non-mode frequencies, the fluctuations are a mix of adiabatic and
non-adiabatic, incoherent fluctuations (of which only the non-adiabatic
ones contribute stochastic work).

\section{Concluding remarks}

We have shown how to decompose the fluid variables into horizontal
averages and fluctuations, and how to use this decomposition to separate
the equations for the radial {\pmode}s (\Eqs{massavL.eq}{energyavL.eq})
from the full equations (\Eqs{8}{8b}).  The equations for the radial
modes are similar to those for a 1-D stratified medium with the
addition of two extra terms -- the gradient of the turbulent pressure
and the gradient of the convective plus kinetic energy fluxes.  There
are two additional subtle differences from the 1-D equations -- the
radiative flux may differ significantly from that of a 1-D model with
the same mean structure and the gas pressure is not the same as the
pressure for the average density and internal energy.

We have used the decomposed fluid equations to derive an expression
for the stochastic excitation rate of the radial {\pmode}s
(\Eq{e11}) in terms of the $PdV$ work of the
non-adiabatic, incoherent, random, convectively produced pressure
fluctuations and the mode compression, and tested the expression
by applying it to a numerical experiment of sufficient length to
measure the mode energy loss directly from the observed mode power
and spectral line width.

The price that had to be paid for obtaining a sufficiently long
duration experiment (43 solar hours in this case) was that the
numerical resolution could not be very large.  From previous
convergence investigations we know that various aspects of the
numerical models depend to a quite varying degree on the numerical
resolution \citep{Stein+Nordlund97a}.  The thermal mean structure, for
example, is very robust, while the peak in relative turbulent
pressure near the surface depends more sensitively on numerical
resolution.

In the subsequent paper \citep[Paper II]{Stein+Nordlund98exc}, we
use higher resolution models to determine the mode excitation power
more accurately, and compare it directly with helioseismic results.
The duration of these experiments are not sufficient to allow the
excited modes to be resolved in frequency, but this is not necessary
to determine the mode excitation power from \Eq{e11}.  In addition,
the high resolution experiments are used to reveal the spatial
locations that contribute most of the excitation power, and to
investigate the nature of the mechanism responsible for the
excitation.

Using the type of formalism put forward in the present paper it is also
possible to analyze the mode physics (intrinsic) contributions to mode
damping and frequency shifts, but this is work for the future.

\acknowledgments

The work of {\AA}N was supported in part by the Danish Research Foundation,
through its establishment of the Theoretical Astrophysics Center.
The work of RFS was supported in part by grants NAG 5-4031 and NAG
5-8053 from NASA and AST 9521785 and AST 9819799 from the US National
Science Foundation.  We thank the High Altitude Observatory / National
Center for Atmospheric Research for hospitality during the writing of
this paper.

\bibliographystyle{apj}
\bibliography{apj-jour,convection,Aake,oscillations}

\begin{thebibliography}{42}
\expandafter\ifx\csname natexlab\endcsname\relax\def\natexlab#1{#1}\fi

\bibitem[{Antia \& Chitre(1996)}]{BOMBAY96}
Antia, H.~M. \& Chitre, S.~M., eds. 1996, Bull. of the Atronomical Soc. of
  India, Vol.~24, Windows on the Sun's Interior

\bibitem[{{Atroshchenko} \& {Gadun}(1994)}]{Atroshchenko+Gadun94}
{Atroshchenko}, I.~N. \& {Gadun}, A.~S. 1994, \aap, 291, 635

\bibitem[{Balmforth(1992{\natexlab{a}})}]{Balmforth92b}
Balmforth, N.~J. 1992{\natexlab{a}}, 255, 632

\bibitem[{Balmforth(1992{\natexlab{b}})}]{Balmforth92c}
---. 1992{\natexlab{b}}, 255, 639

\bibitem[{{Bogdan} {et~al.}(1993){Bogdan}, {Cattaneo}, \&
  {Malagoli}}]{Bogdan+93}
{Bogdan}, T.~J., {Cattaneo}, F., \& {Malagoli}, A. 1993, \apj, 407, 316

\bibitem[{Brown(1993)}]{GONG92}
Brown, T.~M., ed. 1993, Astronomical Society of the Pacific Conference Series,
  Vol.~42, GONG'92: Seismic investigation of the Sun and stars, San Francisco

\bibitem[{{Christensen-Dalsgaard}(1988)}]{JCD88}
{Christensen-Dalsgaard}, J. 1988, in ESA, Seismology of the Sun and Sun-Like
  Stars p 431-450 (SEE N89-25819 19-92), 431--450

\bibitem[{Christensen-Dalsgaard
  {et~al.}(1996{\natexlab{a}})Christensen-Dalsgaard, {D{\"a}ppen}, Ajukov,
  Anderson, Antia, Basu, Baturin, Berthomieu, Chaboyer, Chitre, Cox, Demarque,
  Donatowicz, Dziembowski, Gabriel, Gough, Guenther, Guzik, Harvey, Hill,
  Houdek, Iglesias, Kosovichev, Leibacher, Morel, Proffitt, Provost, Reiter,
  Jr., J., Rogers, Roxburgh, Thompson, \& Ulrich}]{JCD+96science}
Christensen-Dalsgaard, J., {D{\"a}ppen}, W., Ajukov, S.~V., Anderson, E.~R.,
  Antia, H.~M., Basu, S., Baturin, V.~A., Berthomieu, G., Chaboyer, B., Chitre,
  S.~M., Cox, A.~N., Demarque, P., Donatowicz, J., Dziembowski, W.~A., Gabriel,
  M., Gough, D.~O., Guenther, D.~B., Guzik, J.~A., Harvey, J.~W., Hill, F.,
  Houdek, G., Iglesias, C.~A., Kosovichev, A.~G., Leibacher, J.~W., Morel, P.,
  Proffitt, C.~R., Provost, J., Reiter, J., Jr., R., J., E., Rogers, F.~J.,
  Roxburgh, I.~W., Thompson, M.~J., \& Ulrich, R.~K. 1996{\natexlab{a}},
  Science, 272, 1286

\bibitem[{Christensen-Dalsgaard
  {et~al.}(1996{\natexlab{b}})Christensen-Dalsgaard, {D{\"a}ppen}, \& {et
  al.}}]{JCD+ea96science}
Christensen-Dalsgaard, J., {D{\"a}ppen}, W., \& {et al.} 1996{\natexlab{b}},
  Science, 272, 1286

\bibitem[{{Christensen-Dalsgaard} {et~al.}(1988){Christensen-Dalsgaard},
  {Dappen}, \& {Lebreton}}]{JCD+WD+YL88}
{Christensen-Dalsgaard}, J., {Dappen}, W., \& {Lebreton}, Y. 1988, \nat, 336,
  634

\bibitem[{{Georgobiani} {et~al.}(2000){Georgobiani}, {Kosovichev}, {Nigam},
  {Nordlund}, \& {Stein}}]{Georgobiani+99b}
{Georgobiani}, D.~G., {Kosovichev}, A.~G., {Nigam}, R., {Nordlund}, {\AA}., \&
  {Stein}, R.~F. 2000, \apjl, 530, L139

\bibitem[{{Goldreich} \& {Keeley}(1977)}]{Goldreich+Keeley77b}
{Goldreich}, P. \& {Keeley}, D.~A. 1977, \apj, 212, 243

\bibitem[{{Goldreich} \& {Kumar}(1988)}]{Goldreich+Kumar88}
{Goldreich}, P. \& {Kumar}, P. 1988, \apj, 326, 462

\bibitem[{{Goldreich} \& {Kumar}(1990)}]{Goldreich+Kumar90}
---. 1990, \apj, 363, 694

\bibitem[{{Goldreich} {et~al.}(1994){Goldreich}, {Murray}, \&
  {Kumar}}]{Goldreich+94b}
{Goldreich}, P., {Murray}, N., \& {Kumar}, P. 1994, \apj, 424, 466

\bibitem[{Hoeksema {et~al.}(1995)Hoeksema, Domingo, Fleck, \& Battrick}]{SOHO4}
Hoeksema, J.~T., Domingo, V., Fleck, B., \& Battrick, B., eds. 1995, ESA SP,
  Vol. 376, Fourth SOHO Workshop: Helioseismology, ESTEC,Nordwijk

\bibitem[{{Mihalas} {et~al.}(1990){Mihalas}, {Hummer}, {Mihalas}, \&
  {D{\"a}ppen}}]{MHD90}
{Mihalas}, D., {Hummer}, D.~G., {Mihalas}, B.~W., \& {D{\"a}ppen}, W. 1990,
  \apj, 350, 300

\bibitem[{{Musielak} {et~al.}(1994){Musielak}, {Rosner}, {Stein}, \&
  {Ulmschneider}}]{Musielak+94}
{Musielak}, Z.~E., {Rosner}, R., {Stein}, R.~F., \& {Ulmschneider}, P. 1994,
  \apj, 423, 474

\bibitem[{Nordlund(1982)}]{Nordlund82}
Nordlund, {\AA}. 1982, \aap, 107, 1

\bibitem[{Nordlund(1985)}]{Nordlund85f}
---. 1985, Solar Physics, 100, 209

\bibitem[{Nordlund \& Dravins(1990)}]{Nordlund+Dravins90}
Nordlund, {\AA}. \& Dravins, D. 1990, \aap, 228, 155

\bibitem[{Nordlund \& Stein(1991{\natexlab{a}})}]{Nordlund+Stein91b}
Nordlund, {\AA}. \& Stein, R.~F. 1991{\natexlab{a}}, in Stellar Atmospheres:
  Beyond Classical Models, ed. L.~Crivellari, I.~Hubeny, \& D.~G. Hummer
  (Kluwer)

\bibitem[{Nordlund \& Stein(1991{\natexlab{b}})}]{Nordlund+Stein91c}
Nordlund, {\AA}. \& Stein, R.~F. 1991{\natexlab{b}}, in Lecture Notes in
  Physics, Vol. 388, Challenges to Theories of the Structure of Moderate Mass
  Stars, ed. D.~Gough \& J.~Toomre (Springer, Heidelberg), 141--146

\bibitem[{Pijpers {et~al.}(1997)Pijpers, J.Christensen-Dalsgaard, \&
  Rosenthal}]{SCORE96}
Pijpers, F.~P., J.Christensen-Dalsgaard, \& Rosenthal, C.~S., eds. 1997, Solar
  Convection, Oscillations and their Relationship; SCORe'96 (Dordrecht: Kluwer
  Academic Press)

\bibitem[{Roca~Cortes {et~al.}(1999)Roca~Cortes, Montanes, Palle,
  Perez~Hernandez, Jimenez, Regula, \& the GOLF~Team}]{RocaCortes+99}
Roca~Cortes, T., Montanes, P., Palle, P.~L., Perez~Hernandez, F., Jimenez, A.,
  Regula, C., \& the GOLF~Team. 1999, in ASP Conf. Ser., Vol. 173, Theory and
  Tests of Convective Energy Transport, ed. A.~Gimenez, E.~Guinan, \&
  B.~Montesinos, 305

\bibitem[{{Rogers} {et~al.}(1996){Rogers}, {Swenson}, \& {Iglesias}}]{OPAL96}
{Rogers}, F.~J., {Swenson}, F.~J., \& {Iglesias}, C.~A. 1996, \apj, 456, 902

\bibitem[{{Rosenthal} {et~al.}(1998){Rosenthal}, {Christensen-Dalsgaard},
  {Kosovichev}, {Nordlund}, {Reiter}, {Rhodes}, {Schou}, {Stein}, \&
  {Trampedach}}]{Rosenthal+98gong}
{Rosenthal}, C.~S., {Christensen-Dalsgaard}, J., {Kosovichev}, A.~G.,
  {Nordlund}, A.~A., {Reiter}, J., {Rhodes}, E.~J., J., {Schou}, J., {Stein},
  R.~F., \& {Trampedach}, R. 1998, in Structure and Dynamics of the Interior of
  the Sun and Sun-like Stars, SOHO 6/GONG 98 Workshop, E123--+

\bibitem[{Rosenthal {et~al.}(1999)Rosenthal, Christensen-Dalsgaard, Nordlund,
  Stein, \& Trampedach}]{Rosenthal+97a}
Rosenthal, C.~S., Christensen-Dalsgaard, J., Nordlund, {\AA}., Stein, R.~F., \&
  Trampedach, R. 1999, \aap, 351, 689

\bibitem[{Solanki {et~al.}(1996)Solanki, R\"uedi, Bianda, \&
  Steffen}]{Solanki+96}
Solanki, S., R\"uedi, I., Bianda, M., \& Steffen, M. 1996, \aap, 308, 623

\bibitem[{Steffen(1988)}]{Steffen88b}
Steffen, M. 1988, in Advances in Helio- and Asteroseismology, ed.
  J.~Christensen-Dalsgaard \& S.~Frandsen (Reidel, Dordrecht), 379--382

\bibitem[{Steffen \& Freytag(1991)}]{Steffen+Freytag91}
Steffen, M. \& Freytag, B. 1991, in Reviews in Modern Astronomy, ed. G.~Klare,
  Vol.~4 (Springer, Heidelberg), 43--60

\bibitem[{Steffen {et~al.}(1989)Steffen, Ludwig, \& Kr\"uss}]{Steffen+89}
Steffen, M., Ludwig, H.-G., \& Kr\"uss, A. 1989, \aap, 213, 371

\bibitem[{Stein(1967)}]{Stein67}
Stein, R.~F. 1967, Solar Physics, 2, 385

\bibitem[{Stein(1968)}]{Stein68}
---. 1968, \apj, 297, 154

\bibitem[{Stein {et~al.}(1989)Stein, Nordlund, \& {\AA}}]{Stein+Nordlund89a}
Stein, R.~F., Nordlund, \& {\AA}, Kuhn, J. 1989, in Solar and Stellar
  Granulation, ed. R.~Rutten \& G.~Severino (Kluwer Academic Press), 381--399

\bibitem[{Stein \& Nordlund(1989)}]{Stein+Nordlund89b}
Stein, R.~F. \& Nordlund, {\AA}. 1989, \apj, 342, L95

\bibitem[{Stein \& Nordlund(1991)}]{Stein+Nordlund91b}
Stein, R.~F. \& Nordlund, {\AA}. 1991, in Lecture Notes in Physics, Vol. 388,
  Challenges to Theories of the Structure of Moderate Mass Stars, ed. D.~Gough
  \& J.~Toomre (Springer, Heidelberg), 195--212

\bibitem[{Stein \& Nordlund(1994)}]{Stein+Nordlund94a}
Stein, R.~F. \& Nordlund, {\AA}. 1994, in Infrared Solar Physics, IAU Symposium
  154, ed. D.~Rabin, J.~Jefferies, \& C.~Lindsey (Dordrecht: Kluwer), 225--238

\bibitem[{Stein \& Nordlund(1998)}]{Stein+Nordlund97a}
---. 1998, \apj, 499, 914

\bibitem[{Stein \& Nordlund(2000)}]{Stein+Nordlund98exc}
---. 2000, \apj, (submitted)

\bibitem[{Stein {et~al.}(1988)Stein, Nordlund, \& Kuhn}]{Stein+88}
Stein, R.~F., Nordlund, {\AA}., \& Kuhn, J.~R. 1988, in ESA, Vol. SP-286, Proc.
  Symp. Seismology of the Sun and Sun-like Stars, ed. E.~J. Rolfe, 529

\bibitem[{Ulrich {et~al.}(1995)Ulrich, Rhodes~Jr, \& {D{\"a}ppen}}]{GONG94}
Ulrich, R.~K., Rhodes~Jr, E.~J., \& {D{\"a}ppen}, W., eds. 1995, Astronomical
  Society of the Pacific Conference Series, Vol.~76, GONG'94: Helio- and
  Astero-seismology from Earth and Space, San Francisco

\end{thebibliography}

\newpage
\section*{Figure Captions}
\setcounter{figure}{0}

\figcaption[work.ps]
{The excitation power (small pluses) in the
neighborhood of the frequencies of the three radial modes that
are excited to measurable amplitudes in the box.  The average
excitation is shown as diamonds, and the average energy loss per
unit time is shown as squares.  The results are from a numerical
experiment with realistic opacities and equation of state, covering
43 hours of solar time at a resolution of $63\times63\times63$
\citep{Georgobiani+99b}.
Note that, because of the relatively low resolution required
for such a long run to be affordable, the results are not directly
comparable to solar values.  The mode mass of the box also differs
significantly from the solar mode mass at the same frequency.}

\figcaption[modes.ps]
{The velocity mode shapes for the three
radial oscillations, at 2.5, 3.9, and 5.6 mHz, that are observed in
the 43 hour simulation.  For comparison, the corresponding modes
from Model S of \citet{JCD+96science}, normalized to the same mode
mass within the box, are also shown (plus symbols) for the two modes
whose frequencies are below the acoustic cut-off frequency.}

\figcaption[Pna.ps]
{The power-spectrum of the non-adiabatic pressure
fluctuations, \Eq{w7a}, in a layer close to the surface of the 43 hour
simulation.}

\end{document}